\documentclass[twoside]{ilcws07}
\usepackage[latin1]{inputenc}
\usepackage[dvips]{graphicx,epsfig,color}
\usepackage{wrapfig,rotating}
\usepackage{amssymb,amsmath,array}

\begin{document}
\title{
Polarization-independent CP-odd Observable in $e^+e^-$ Chargino
Production at One Loop} 
\author{Per Osland$^1$, Jan Kalinowski$^2$, Krzysztof Rolbiecki$^2$
and Alexander Vereshagin$^1$
\vspace{.3cm}\\
1- Department of Physics and Technology - University of Bergen \\
Postboks 7803, N-5020 Bergen - Norway
\vspace{.1cm}\\
2- Institute of Theoretical Physics - University of Warsaw\\
PL-00681 Warsaw - Poland\\
}

\maketitle

\begin{abstract}
We discuss CP violation in the process $e^+e^- \rightarrow
\tilde\chi^+_i\tilde\chi^-_j$ with unpolarized beams.  When the scalars are
heavy, the box-diagram results constitute a major part of the full result.
However, there are situations when the vertex and self-energy corrections
dominate over the box diagrams. We also comment on CP violation in the final
chargino decay.
\end{abstract}

\section{Introduction}

We review recent work on CP violation in unpolarized $e^+e^-
\rightarrow \tilde\chi^+_i\tilde\chi^-_j$
\cite{url,Osland:2007xw,Kalinowski}. Let us consider the process with
unpolarized initial beams:
\begin{equation}
e^+(p_1) + e^-(p_2) \to \tilde\chi_i^+(k_1) + \tilde\chi_j^-(k_2).
\label{eq:original-process}
\end{equation}
The crucial point here is that for $i \neq j$ the charginos do not
form a particle-antiparticle pair. Hence, while the initial state
is in the c.m.\ frame odd under charge conjugation, the final state
has no such symmetry. This leads to the CP-violating effect we
discuss here.

\section{CP violating observable}

Under CP conjugation the $S$-matrix element $\langle \tilde\chi^+_i
({\boldsymbol k}_1), \tilde\chi^-_j ({\boldsymbol k}_2) | S | e^+
({\boldsymbol p}_1), e^- ({\boldsymbol p_2}) \rangle$ of the process
(\ref{eq:original-process}) gets transformed into (up to a phase
which is irrelevant for us):
\begin{alignat}{2}
&\text{CP:} &\quad &\langle \tilde\chi^+_j ( -{\boldsymbol k}_2),
\tilde\chi^-_i ( -{\boldsymbol k}_1) | S | e^+ (- {\boldsymbol
p}_2), e^- ( - {\boldsymbol p}_1) \rangle,
\end{alignat}
which amounts to the following change in the cross section%
\footnote{Of course, the coupling constants at vertices with
charginos should be considered as functions of the chargino masses
$m_i, m_j$, or, better, the mass indices $i,j$. }: ${\boldsymbol
p}_1 \leftrightarrow -{\boldsymbol p}_2, \; {\boldsymbol k}_1
\leftrightarrow -{\boldsymbol k}_2, \; m_i \leftrightarrow m_j$. Due
to Poincar\'{e} invariance the {\em unpolarized} cross section
$d\sigma_0$ may depend only on the masses $m_i, m_j$ and on two
independent scalar variables, say, on Mandelstam's \mbox{$s \equiv
(p_1 + p_2)^2$} and \mbox{$t \equiv (p_1 - k_1)^2$} which obviously
do not change under C or P. Hence, if one sticks to the unpolarized
part only, the CP transformation can be reduced to final-state
chargino mass interchange: $m_i \leftrightarrow m_j$. Therefore, for
equal-mass fermions in the final state
($i = j$) the unpolarized cross section is always P-, C- and CP-even%
\footnote{
The famous forward-backward asymmetry term in the
{\em unpolarized} cross-section of, say,
$e^+e^- \to \mu^+\mu^-$ scattering, which is often referred to as
parity violating, in fact only indicates the presence of a parity
violating term in the interaction, the unpolarized cross-section
itself being, of course, P-even.
}.
In contrast, if the chargino species are different, CP-violating
terms can arise even in the unpolarized cross-section. That is the
effect we will consider, so unless otherwise stated the
final-state chargino masses are taken non-equal. The
polarization-dependent CP-violating observables at one-loop order
require more involved analysis and will not be discussed here.

Calculations show that the tree-level cross section (polarized and
unpolarized) of the process (\ref{eq:original-process}) is CP even
\cite{Bartl:2004xy}, but CP-odd terms do arise in the one-loop
contributions. Therefore, a natural experimental observable to
consider is the ratio
\begin{equation}
\frac{d\sigma_0^{\rm odd}}{d\sigma_0}\ ,
\label{eq:obvservable-definition}
\end{equation}
where
$d\sigma^{\rm odd}$ is the CP-odd part of the corresponding
cross-section:
\begin{equation}
d\sigma_0^{\rm odd} =
\frac{1}{2} \Bigl[ d\sigma_0 - d\sigma_0^{\rm CP} \Bigr] , \quad
d\sigma_0^{\rm CP}
\equiv d\sigma_0 \Bigr|_{m_i \leftrightarrow m_j} .
\label{eq:sigma_odd_def}
\end{equation}

As just mentioned, the CP violation first enters at one loop, thus,
to estimate the effect one should calculate $d\sigma_0^{\rm odd}$ at
the one-loop level. On the other hand, in most of the kinematical
regions far from any resonance, one can expect (see, e.g.\
\cite{Blank:2000uc,Diaz:2002rr,Fritzsche:2004nf,Oller:2005xg}) that
the tree level gives a reasonable approximation to $d\sigma_0$ in
the denominator of Eq.~(\ref{eq:obvservable-definition}). So, we
will deal only with the ratio
\begin{equation}
A_{\rm CP}=\frac{\left. d\sigma_0^{\rm odd} \right|_{\rm 1\; loop}
}{
\left. d\sigma_0 \right|_{\rm tree}}\ .
\label{eq:obvservable-loop/tree}
\end{equation}

\section{Box diagrams vs.\ full one loop contribution}

\begin{wrapfigure}{r}{0.5\columnwidth}
\centerline{\includegraphics[width=0.45\columnwidth]{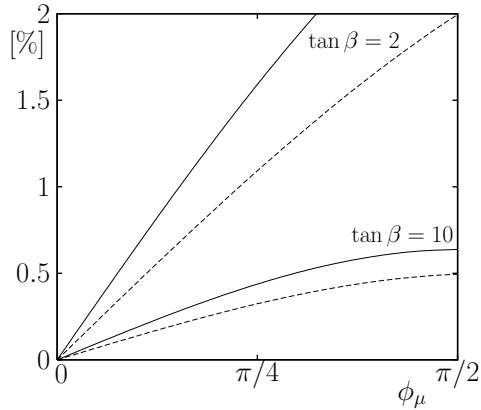}}
\caption{Box-only contribution (dashed lines) vs.\ full one-loop result (full
lines) in the heavy sfermions limit for $\tan\beta=2$ and 10.}
\label{Fig:full_vs_box}
\end{wrapfigure}
In \cite{Osland:2007xw} partial one-loop calculations were provided
for the case of complex higgsino mass parameter $\mu$. A limitation
of that analysis was that it was performed in the heavy slepton
limit, and furthermore, all one-loop triangle vertex corrections to
(\ref{eq:obvservable-loop/tree}) were dropped. The observed effect
turned out to be of the order of a couple of percent\footnote{A
factor of four was lost in the calculation.}, depending on the
chosen MSSM parameters and the kinematics. As explained in
\cite{Osland:2007xw}, this calculation was done just to make sure
that the observable does not vanish, while its magnitude should be
estimated from complete one-loop results.

A full calculation has recently been performed~\cite{Kalinowski}.
The full result turns out to be of the same order as the box-only
estimates. In Fig.~\ref{Fig:full_vs_box} the ``box-only'' and full
one-loop values of the observable (\ref{eq:obvservable-loop/tree})
are plotted as functions of the higgsino phase $\phi_\mu$ for
$\tan\beta = 2$ and $10$.  The other parameters are taken as:
$\sqrt{s} = 600\ {\rm GeV}$, the polar scattering angle $\theta =
\pi/3$, the Higgsino mass parameter $\mu = |\mu|e^{i\phi_\mu}$,
$|\mu| = 300\ {\rm GeV}$, the SU(2) gaugino mass parameter $M_2 =
200\ {\rm GeV}$, the U(1) gaugino mass parameter (taken to be real)
$M_1 = 250\ {\rm GeV}$. The common SUSY breaking mass of the scalars
(for the full one-loop calculation) is 1~TeV.

\begin{figure}
\label{Fig:selectron box contribution}
\begin{center}
\includegraphics[width=0.40\columnwidth]{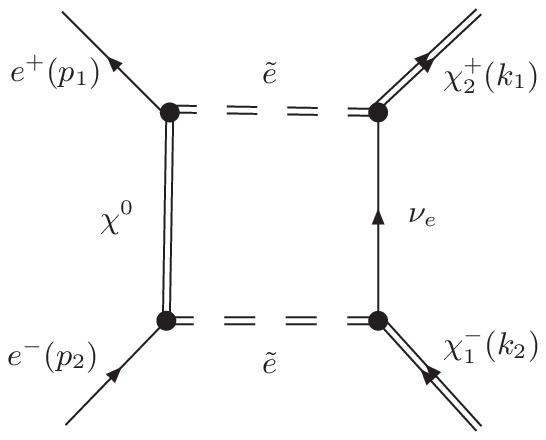}
\hskip0.4cm
\includegraphics[width=0.49\columnwidth]{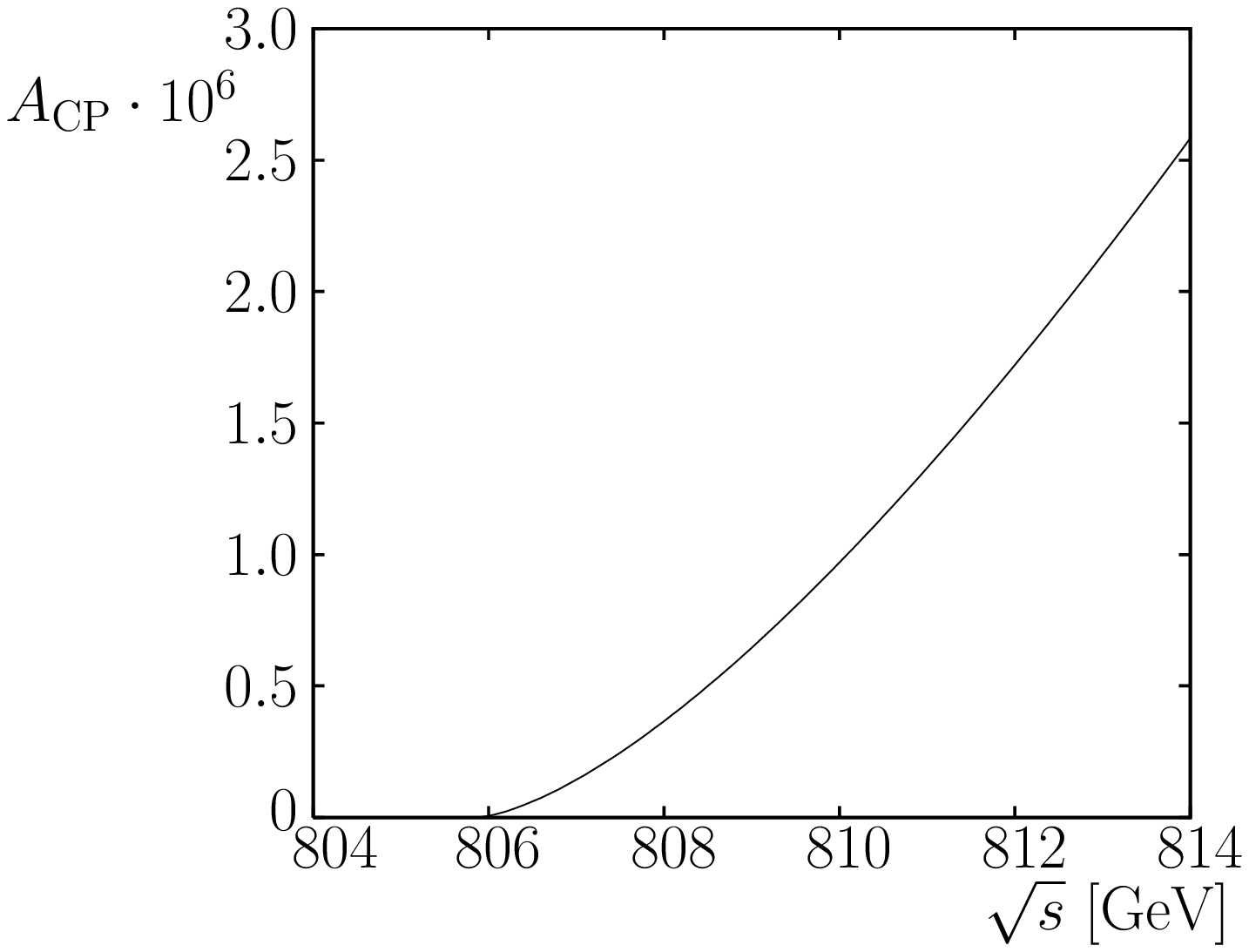}
\caption{Selectron exchange box diagram and its contribution to
(\ref{eq:obvservable-loop/tree}). The selectron mass is 403~GeV.}
\end{center}
\end{figure}

The qualitative agreement for the gauge box contribution alone can
be (at least, partially) explained. Indeed, a closer look at the
expression for the $Z$-boson exchange contribution (Eq.~(4.1) in
\cite{Osland:2007xw} gives the $D$-function part) shows, that only
the imaginary part of the box integral can affect the observable.
Since in the heavy slepton limit the position of the threshold
singularity is high, the integral remains real in the kinematical
region we consider. The selectron exchange box diagram provides a
nice illustration: when we raise the c.m.\ energy above the
selectron pair production threshold the selectron box diagram
develops an absorptive part and its contribution to the asymmetry
(\ref{eq:obvservable-loop/tree}) is non-zero, see
Fig.~\ref{Fig:selectron box contribution} (the selectron mass is
$403$~GeV).  Similar statements can be made about most of the
diagrams contributing to (\ref{eq:obvservable-loop/tree}) at the
one-loop order.

The above argument also indicates that in a scenario with lighter sparticles,
other diagrams with vertex and self-energy corrections cannot be neglected, as
demonstrated in~\cite{Kalinowski}. It was also shown there that for the case
of CP-violating origin in the top squark sector the box diagrams do not
contribute and the CP asymmetry receives contributions only from vertex and
self-energy diagrams.

\section{Chargino decay: interference with CP violating effects}

Since charginos are not stable particles and decay finally to
leptons/quarks and the LSP, in a realistic experiment one has to
take into account also chargino decays. On the other hand we know
that also in chargino decays it is possible to obtain CP-violating
effects at one-loop level~\cite{Eberl:2005ay}. Therefore one can
worry if CP-violating effects in the decay would not cancel similar
effects in the chargino production. However, a consideration similar
to one presented in~\cite{Osland:2007xw} helps here. As shown in
that paper, at the one-loop level the observable
(\ref{eq:obvservable-loop/tree}) among other pieces contains the
$D$-function integral.

To cancel such a contribution at any kinematical point, one needs a
corresponding contribution from the final particle decay. So, the
only way is the box diagram (e.g.\ the $Z$-exchange box---see
\cite{Osland:2007xw}, Fig.~2) attached to one of the external legs.
Even if the mass splitting between charginos is larger than $2 m_Z$,
the kinematic configuration of the box diagram in the decay is
completely different from the one in the production, so the
cancelation of different CP-odd contributions is in general not
possible. This statement becomes trivial if the mass splitting is
smaller than $2 m_Z$ and no CP asymmetry arises in the decay due to
double $Z$ exchange diagram. Moreover it is even possible to arrange
parameters in such a way that no 2-body decay channels remain open
for charginos and therefore no CP-odd contribution due to chargino
decays enter in the full production+decay process, but still
allowing for such contributions in the chargino production.
Therefore we conclude that in general CP-odd effects in the
production process can not be canceled by CP-odd effects in the
decays of charginos.

\section{Summary}
We have demonstrated that the CP asymmetry built from unpolarized
cross sections for non-diagonal chargino pairs in $e^+e^-$
annihilation can arise at the one-loop order.

\section*{Acknowledgements}
JK and KR have been supported by the Polish Ministry of Science and
Higher Education Grant No.\ 1 P03B 108 30. This research has been
supported in part by the Research Council of Norway.


\begin{footnotesize}


\end{footnotesize}



\begin{thebibliography}{99}
\bibitem{url} Slides: \\
\verb$http://ilcagenda.linearcollider.org/contributionDisplay.py?contribId=56&sessionId=69&confId=1296$

\bibitem{Osland:2007xw}
  P.~Osland and A.~Vereshagin,
  Phys.\ Rev.\  D {\bf 76}, 036001 (2007)
  [arXiv:0704.2165 [hep-ph]].

\bibitem{Kalinowski}
K. Rolbiecki and J. Kalinowski, arXiv:0709.2994 [hep-ph].

\bibitem{Bartl:2004xy}
S.~Y.~Choi, A.~Djouadi, H.~S.~Song and P.~M.~Zerwas,
  Eur.\ Phys.\ J.\  C {\bf 8}, 669 (1999)
  [arXiv:hep-ph/9812236];\\
S.~Y.~Choi, A.~Djouadi, M.~Guchait, J.~Kalinowski, H.~S.~Song and P.~M.~Zerwas,
  Eur.\ Phys.\ J.\  C {\bf 14}, 535 (2000)
  [arXiv:hep-ph/0002033];\\
A.~Bartl, K.~Hohenwarter-Sodek, T.~Kernreiter and H.~Rud,
  Eur.\ Phys.\ J.\ C {\bf 36}, 515 (2004)
  [arXiv:hep-ph/0403265].

\bibitem{Blank:2000uc}
  T.~Blank and W.~Hollik,
  in {\em 2nd ECFA/DESY Study 1998--2001};
  arXiv:hep-ph/0011092.

\bibitem{Diaz:2002rr}
  M.~A.~Diaz and D.~A.~Ross,
  JHEP {\bf 0106}, 001 (2001)
  [arXiv:hep-ph/0103309].

\bibitem{Fritzsche:2004nf}
  T.~Fritzsche and W.~Hollik,
  Nucl.\ Phys.\ Proc.\ Suppl.\  {\bf 135}, 102 (2004)
  [arXiv:hep-ph/0407095].

\bibitem{Oller:2005xg}
  W.~Oller, H.~Eberl and W.~Majerotto,
  Phys.\ Rev.\  D {\bf 71}, 115002 (2005)
  [arXiv:hep-ph/0504109].

\bibitem{Eberl:2005ay}
  H.~Eberl, T.~Gajdosik, W.~Majerotto and B.~Schrausser,
  Phys.\ Lett.\  B {\bf 618}, 171 (2005)
  [arXiv:hep-ph/0502112].

\end{thebibliography}
\end{document}